\documentclass[12pt]{article}
 \usepackage{epsfig}
 \usepackage{color}
 \pdfoutput=1
 \def\aprle{\buildrel < \over {_{\sim}}} 
 \def\aprge{\buildrel > \over {_{\sim}}} 

\begin{document}       

 \title{Remarks  on the forces generated by two--neutrino exchange}

 \author{Maurizio Lusignoli  and Silvano Petrarca \\\\
 {\normalsize
 Sapienza, Universit\`{a} di Roma, and INFN, Sezione di Roma}
\\
\normalsize
 Piazza A. Moro 2, I-00185 Roma, Italy 
}

 \date{}          
 \maketitle

\begin{abstract}
A brief up--to--date review of the long range forces generated by two neutrino exchange is presented. The potential due to exchange of a massive neutrino-antineutrino pair between particles carrying  weak charge might be larger than expected if the neutrinos have not only masses but also magnetic moments close to the present experimental bounds. It still remains too small to be observable. 
\end{abstract}

\section{Introduction}
\qquad Many years ago Feinberg and Sucher \cite{Feinberg} calculated the long range part of the potential due to the exchange of a pair of massless neutrinos between particles carrying a weak charge. Later, the same problem was attacked by Hsu and Sikivie \cite{Sikivie}  with a slightly different technique and the same results. The potential decreases as the fifth power of the distance, and is too small to be observed, being proportional to the square of the Fermi constant $G_F$. 

Nowadays we know that neutrinos do have nonvanishing masses, therefore the potential will not be long-range anymore, and it will be exponentially suppressed at distances $r \aprge 1/(2 m_{\nu})$, making its detection even harder. We will further discuss this point.  

It has been suggested that neutrinos may have a nonzero magnetic moment (if their mass is of Dirac type): for a good review, see \cite{Giunti:2008ve}. 
The present limit from particle physics \cite{GEMMA,MUNU,TEXONO} is $\sim 10^{-11} \mu_{\rm Bohr}$ and even tighter limits come from astrophysical processes \cite{Raffelt:1989de}. In the simplest extension of the standard model (adding a singlet, right-handed neutrino for each family) the magnetic moment comes out  much smaller, 3 $10^{-19} m_{\nu}/1 {\rm eV}$ \cite{mu_nu_Dirac}, and several theoretical models 
have been put forward to justify a value closer to the experimental bound without having at the same time too large a neutrino mass \cite{mu_nu_th}. Arguments based on naturalness suggest more stringent bounds for the magnetic moments \cite{natur}. If the magnetic moment is close to the present experimental upper bound, we show in this paper that a contribution to the potential between two electrons due to an exchange of a photon and the pair of neutrinos may be appreciably larger than the pure weak contribution. This surprising result may be easily understood on qualitative grounds, since an interaction with the neutrino magnetic moment should end up in a contribution to the potential proportional to $G_F e~m_{\nu} \mu_{\nu}  =  2\pi~G_F\alpha~ {m_{\nu} \over m_e} ~  {\mu_{\nu} \over \mu_{\rm Bohr}}$ and decreasing as the third power of the distance (over and above the exponential suppression ~ $\exp (- 2m_{\nu}r)$). Therefore  the ratio of this contribution to the pure weak one, barring multiplicative constants, should be of the order of  ${1 \over G_F } \alpha\,{m_{\nu} \over m_e} \, {\mu_{\nu} \over \mu_{\rm Bohr}}\; r^2 $, which for $m_{\nu} \sim 0.05\; {\rm eV}$,  $\mu_{\nu}  \sim  10^{-11}  \mu_{\rm Bohr}$ and at a distance $r \sim 1 \mu$m is about  15000. Moreover, the neutrino magnetic moment induces a further contribution to the vacuum polarization, with an even larger contribution to the potential. In the following sections  we derive the precise expressions for these contributions and discuss the possible effects of neutrino mixing and different neutrino masses. 

This result is somehow only a curiosity, since for neutral atoms the net contribution of the  terms dependent on the magnetic moment vanishes, due to cancellation of contributions coming from oppositely charged particles. 

The paper is organized as follows: in Section 2 we extend the calculation by Feinberg and Sucher \cite{Feinberg} to the case of a neutrino with mass. In Section 3 we perform the calculation of the contribution due to the neutrino magnetic moment. Section 4 is devoted to a discussion of the modifications that are needed, should the difference in neutrino absolute masses be relevant. Section 5 contains our conclusions.

\section{Potential due to massive neutrino exchange}
\label{mass}
\qquad For the sake of simplicity and clarity we will consider first  the potential due to the exchange of a massive neutrino-antineutrino pair between two electrons, as if the electron neutrino was a mass eigenstate. As we will see, this is not too different from a real case of interest. We will furthermore 
consider only the contribution of the weak charged currents, correcting later for the effect of $Z^0$ exchange.  
\begin{figure}[hb]
 \centering
   \includegraphics[width=0.6 \textwidth ]{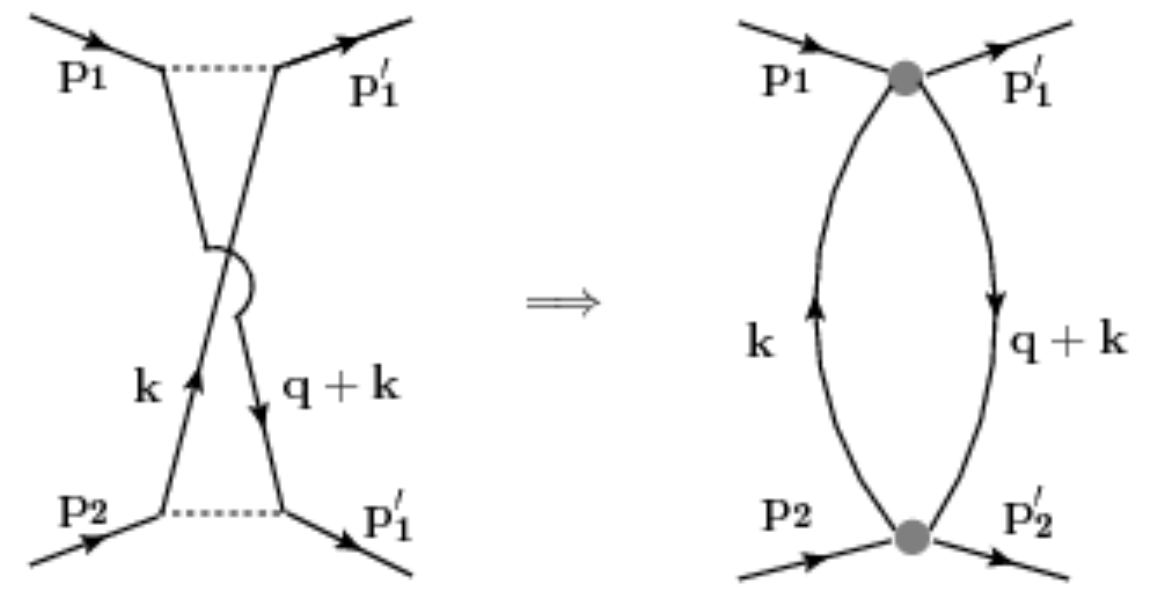}
    \caption{ Charged current two--neutrino exchange between electrons and its Fermi limit.}
   \label{fig:fig1}
   \end{figure}
The scattering amplitude can be written, in the low-energy  (Fermi) limit (see Fig.{\ref {fig:fig1}})
\begin{equation}
-i M = (-i  {G_F \over {\sqrt 2}})^2 H_{\mu \nu} K^{\mu \nu}
\label{ampl}
\end{equation}  
where
\begin{equation}
 H_{\mu \nu} = \bar{u}({p'}_1) \gamma_\mu (a - b \gamma_5) u(p_1) \:
  \bar{u}({p'}_2) \gamma_\nu (a - b \gamma_5) u(p_2)
\label{cov}
\end{equation}
and  
\begin{equation}
 K^{\mu \nu} =  \int {d^4 k \over (2 \pi)^4} {{\rm Tr} [(\widehat{k} +m_{\nu}) \gamma^\mu (1- \gamma_5)
 (\widehat{q}+\widehat{k} +m_{\nu}) \gamma^\nu (1-\gamma_5)] \over {(k^2-m_{\nu}^2)\,((q+k)^2-m_{\nu}^2)}}
\label{loop}
\end{equation}
In eq.(\ref{cov}) $a = b =1$ if only charged weak currents are considered. In eq.(\ref{loop}) $q = p_1 - p'_1$ is the four momentum  transfer. 

 The (divergent) function $K^{\mu \nu}$ has the general form 
\begin{equation}
 K^{\mu \nu} =  A(q^2) \; q^2 g^{\mu \nu} + B(q^2)\; q^{\mu} q^{\nu}\; ,
\label{form}
\end{equation}
and the coefficients $A$ and $B$ are analytic functions of $q^2$, with a branch cut starting at $q^2 = 4 m_\nu^2$, where $m_\nu$ is the neutrino mass. The discontinuity across the cut can be easily evaluated with the Cutkosky rules putting the neutrino lines on the mass shell (i.e. replacing the denominators by delta functions), with the result
\begin{eqnarray}
[ K^{\mu \nu} ] &=&{1 \over 3\pi} \left(1-{4m_{\nu}^2 \over q^2}\right)^{1/2} \left((1+{2 m_{\nu}^2 \over q^2})\;(-q^2 g^{\mu \nu} + q^{\mu} q^{\nu})+3\;m_{\nu}^2 \, g^{\mu \nu} \right) \theta(q^2-4m_{\nu}^2) \nonumber \\
&=&{1 \over 3\pi} \left(1-{4m_{\nu}^2 \over q^2}\right)^{1/2} \left(-\, (q^2-m_{\nu}^2)\, \,g^{\mu \nu} + (1+{2 m_{\nu}^2 \over q^2})\; q^{\mu} q^{\nu}\right)\;\theta(q^2-4m_{\nu}^2)\; .
\label{form2}
\end{eqnarray}
This differs from the result of the analogous calculation of one--loop vacuum polarization in QED and reflects the non--conservation of axial weak current due to the neutrino mass in the last term in the first line of eq.(\ref{form2}).

Following \cite{Feinberg} the long range part of the potential is determined by the equation
\begin{equation}
 V(r) = {1 \over 4 \pi^2 r} \int_{4 m_{\nu}^2}^\infty \rho(q^2) \; e^{-r \sqrt {q^2}} dq^2\;,
\end{equation}
with the spectral function $\rho(q^2)$ given by the discontinuity of the Feynman amplitude divided by $2 i$. 

We are interested in the longest range, spin independent part of the potential, in the limit of nonrelativistic motion of the electrons. In this case the covariant $H_{\mu \nu} $ is dominated by the vector current contribution and it reduces to $u^\dagger ({p'}_1) u(p_1) \: u^\dagger ({p'}_2) u(p_2)\: \delta_{\mu}^0 \delta_{\nu}^0 $. It can be easily seen that the contribution of the term proportional to $q^\mu q^\nu$ is also suppressed in this limit, so that the spectral function is given by the discontinuity of the $A$ term in eq.(\ref{form}), namely
$$\rho(q^2) = {G_F^2 \over 12 \pi} (q^2 - m_{\nu}^2) \left(1-{4m_{\nu}^2 \over q^2}\right)^{1/2} \theta(q^2-4m_{\nu}^2)\; .$$ 
Therefore,
\begin{equation}
 V(r) = {G_F^2 \over 24\, \pi^3\, r^5} \;  \int_{2 m_{\nu} r}^\infty (y^2-m_{\nu}^2\, r^2)\, 
 \sqrt {y^2-4\, m_{\nu}^2\,r^2}\;
 e^{-y}\;dy\;,
 \label{potential}
\end{equation}
which clearly shows that the effect of the neutrino mass is a {\it decrease} of the potential with respect to the massless case for any distance, and not only (as it is obvious) for $2 m_\nu  r \aprge 1$.

The effect of neutral currents would be to add three more diagrams, with $Z^0$ exchanged in one or both of the vertices, and therefore to modify the parameters in the definition of eq.(\ref{cov}) to $a= 1/2 + 2 \sin^2 \theta_W$ and $b=1/2$ if only $\nu_e$ is exchanged in the loops. However the possibility of having neutrinos of different flavours (although for the moment with the same mass) in the $Z^0$ mediated graph leads to two further contributions  to $H_{\mu \nu}$ with $a= -1/2 + 2 \sin^2 \theta_W$ and $b=-1/2$. The dominant vector current contribution would remain as given in eq.(\ref{potential}) if  $\sin^2 \theta_W  = 1/4$, and the value of the weak angle does not differ much from this number.

\section{Effect of a magnetic moment}
\label{munu}
\qquad We now consider the possibility  that the electron neutrino has a Dirac mass $m_{\nu}$ and a magnetic moment $\mu_{\nu}$. In this case, further diagrams are possible, with photon exchange as in Fig.({\ref{fig:fig2}a), coupled to the neutrino magnetic moment. The contribution to the Feynman amplitude is
\begin{equation}
-i M' =  i  {G_F \over {\sqrt 2}} \,e\,\mu_{\nu} H'_{\mu \nu} K'^{\mu \nu}\; ,
\label{muampl}
\end{equation}  
where now
$$H'_{\mu \nu} = \bar{u}({p'}_1) \gamma_\mu  u(p_1) \:
  \bar{u}({p'}_2) \gamma_\nu (a - b \gamma_5) u(p_2)$$
($a=b=1$ if only charged currents are considered) and
\begin{eqnarray}
 K'^{\mu \nu} &=&  {q_{\rho} \over q^2} \int {d^4 k \over (2 \pi)^4} {{\rm Tr} [(\widehat{k} +m_{\nu}) 
 \sigma^{\mu \rho}
 (\widehat{q}+\widehat{k} +m_{\nu}) \gamma^\nu (1-\gamma_5)] \over {(k^2-m_{\nu}^2)\,((q+k)^2-m_{\nu}^2)}} =  \nonumber \\
&=& 4\,i\,m_{\nu} \left(g^{\mu\nu}- {q^\mu q^\nu \over q^2}\right) \int {d^4 k \over (2 \pi)^4}
 {1 \over {(k^2-m_{\nu}^2)\,((q+k)^2-m_{\nu}^2)}} + ...
 \label{Kprime}
 \end{eqnarray}
 In eq.(\ref{Kprime}) we have omitted a term that gives no contribution to the discontinuity. 
Notice that the magnetic moment interaction flips the helicity, and therefore this contribution is proportional to the neutrino mass $m_\nu$. We also note that the term proportional to 
$q^\mu q^\nu$ gives a vanishing contribution, when dotted with $H'_{\mu \nu}$.
 \begin{figure}[ht]
 \centering
   \includegraphics[width=0.6 \textwidth ]{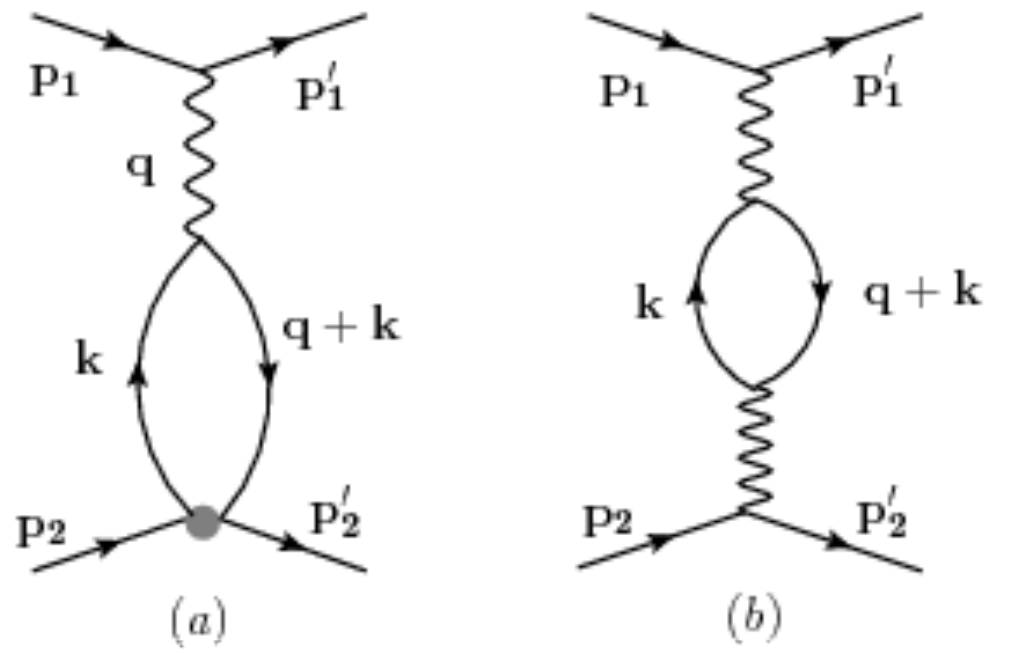}
    \caption{ Feynman graphs for the neutrino magnetic moment contributions to  two--neutrino exchange between electrons. }
   \label{fig:fig2}
   \end{figure}

 Proceeding as above and taking the absorptive part, we obtain
 \begin{equation}
[ K'^{\mu \nu} ] ={- i \over 2 \pi}\; m_\nu \left(1-{4m_{\nu}^2 \over q^2}\right)^{1/2}\;\left(g^{\mu \nu}
- {q^\mu q^\nu \over q^2} \right)\;\theta(q^2-4m_\nu^2)
\label{absKprime}
\end{equation}
and the spectral function
$$
\rho'(q^2) =  {G_F \over 4\sqrt{2} \pi} e m_\nu \mu_\nu \left(1-{4m_{\nu}^2 \over q^2}\right)^{1/2}
\;\theta(q^2-4m_\nu^2)\;.
$$ 
The contribution to the spin independent, longest range part of the potential is therefore
\begin{eqnarray}
 V'(r) &=&  {1 \over 4 \pi^2 r}\quad  \int_{4 m_{\nu}^2}^\infty \rho'(q^2) \; e^{-r \sqrt {q^2}} dq^2 \; = 
 \nonumber \\
 &=&  {{G_F\;\alpha} \over {4 \sqrt{2}\, \pi^2 \, r^3}} \; {m_\nu \over m_e}\; 
 {\mu_\nu \over \mu_{\rm Bohr}}\; \int^\infty_{2 m_\nu r} \sqrt{y^2-4\,m_\nu^2\,r^2}\;e^{-y}\;dy\; ,
 \label{potprime}
 \end{eqnarray}
 where we have introduced the Bohr magneton $ \mu_{\rm Bohr}$, the fine structure constant 
 $\alpha$ and the electron mass $m_e$. The sign of this term depends on the relative orientation between the neutrino spin and magnetic moment (sign of $\mu_{\nu}$).
 
 Again, the effect of neutral currents is simply to add a similar contribution, but with 
 $a=1/2-2\sin^2\theta_W$ and $b=1/2$ (an overall minus sign is due to a Fierz rearrangement made in the previous calculation). If  neutrinos of different flavor have also magnetic moments, they should contribute analogous neutral current terms. 
 Also, it is clear that another contribution is obtained exchanging the r\^ole of electrons, and the final result for the longest range, spin independent part must therefore be doubled.
 
 If neutrinos have nonzero magnetic moments, also the diagram in Fig.({\ref{fig:fig2}b) with only electromagnetic interactions  gives a contribution, that is proportional to the square of the very small number $\mu_\nu / \mu_{\rm Bohr} \aprle 10^{-11}$ and has the square of the fine structure constant $\alpha$  and a factor 1/$m_e^2$ instead of the Fermi constant $G_F$. 
 Moreover, it does not vanish even for massless neutrinos. The calculation of this contribution to the longest range, spin independent part of the potential gives the result:
 \begin{equation}
 V''(r) =  {1 \over 12 \pi r^3}\; {\alpha^2 \over m_e^2}\; \left({\mu_\nu \over \mu_{\rm Bohr}}\right)^2 \; \int^\infty_{2 m_\nu r} \left(1+{8\,m_\nu^2\,r^2 \over y^2}\right) \,
 \sqrt{y^2-4\,m_\nu^2\,r^2}\;e^{-y}\;dy  
 \label{potsecond}
 \end{equation}
 which for massless neutrinos reduces to
 $$
V''(r) =  {1 \over 12 \pi r^3} \;{\alpha^2 \over m_e^2}\; \left({\mu_\nu \over \mu_{\rm Bohr}}\right)^2\;.
$$

The effects of magnetic moment  shown in eqs.(\ref{potprime}, \ref{potsecond}) decrease
 as the third power of the distance, and are therefore able to overcome the potential in eq.(\ref{potential}) that decreases as $r^{-5}$. The term $V''(r)$ is a further contribution to the vacuum polarization, that in QED gives rise to the so--called Uehling potential \cite{Ueh}:
 \begin{eqnarray}
 V^{QED}(r) &=&  {2 \alpha^2  \over 3 \pi r}\;  
  \int^\infty_{2 m_e r} {1 \over y^2}\; \left(1+{2\,m_e^2\,r^2 \over y^2}\right) \,
 \sqrt{y^2-4\,m_e^2\,r^2}\;e^{-y}\;dy  
 \label{potUeh}
 \end{eqnarray}
 This potential is exponentially suppressed for distances larger than half electron Compton wavelength (0.19 pm) and therefore there is a wide range of distances where the neutrino contribution would be more important, if neutrinos have a magnetic moment. It is obvious anyhow that the neutrino term is totally negligible with respect to the zero--order term  (the Coulomb potential) given the smallness of its coefficient. 
 
 A comparison among the neutrino--exchange potentials follows. If  the magnetic moment is larger than  $10^{-15}  \mu_{\rm Bohr}$ one has $V''(r) > |V'(r)|$ for any $r$, given the present experimental upper bound on the neutrino mass \footnote{In the minimally extended standard model with Dirac neutrinos, the magnetic moment is proportional to the mass with a very small coefficient \cite{mu_nu_Dirac}. In this case the term $|V'|$ would be larger than $V''$, and both would be smaller than $V$.}. 
 For $\mu_\nu \sim 10^{-11} \mu_{\rm Bohr}$ the term $V''(r)$ dominates over $V(r)$ for  $r \aprge$ 9$\cdot 10^{-12}$m for any possible value of the neutrino mass. The comparison between the terms $|V'(r)|$ and $V(r)$ shows that $|V'|$ dominates for distances larger than $\sim$ 1.2 (5.5, 13)$\cdot 10^{-9}$m for $m_{\nu}$ = 1 (0.05, 0.0087) eV (i.e the kinematical limit, the value obtained from atmospheric and long--baseline experiments and the value deduced from solar and Kamland results).
 
To illustrate the above points, we report in Fig.({\ref{fig:fig3}) the potentials (in eV) given in eqs.(\ref{potential}, \ref{potprime}, \ref{potsecond}) and  the potential $V_{FS}(r)$ for the massless neutrino case (first derived in \cite{Feinberg}) as  functions of the distance $r$ (in $\mu$m).  The neutrino mass and magnetic moment are assumed equal to 1 eV and 10$^{-11}$ Bohr magnetons, respectively.  
 \begin{figure}[h]
 \centering
   \includegraphics[width=0.8 \textwidth ]{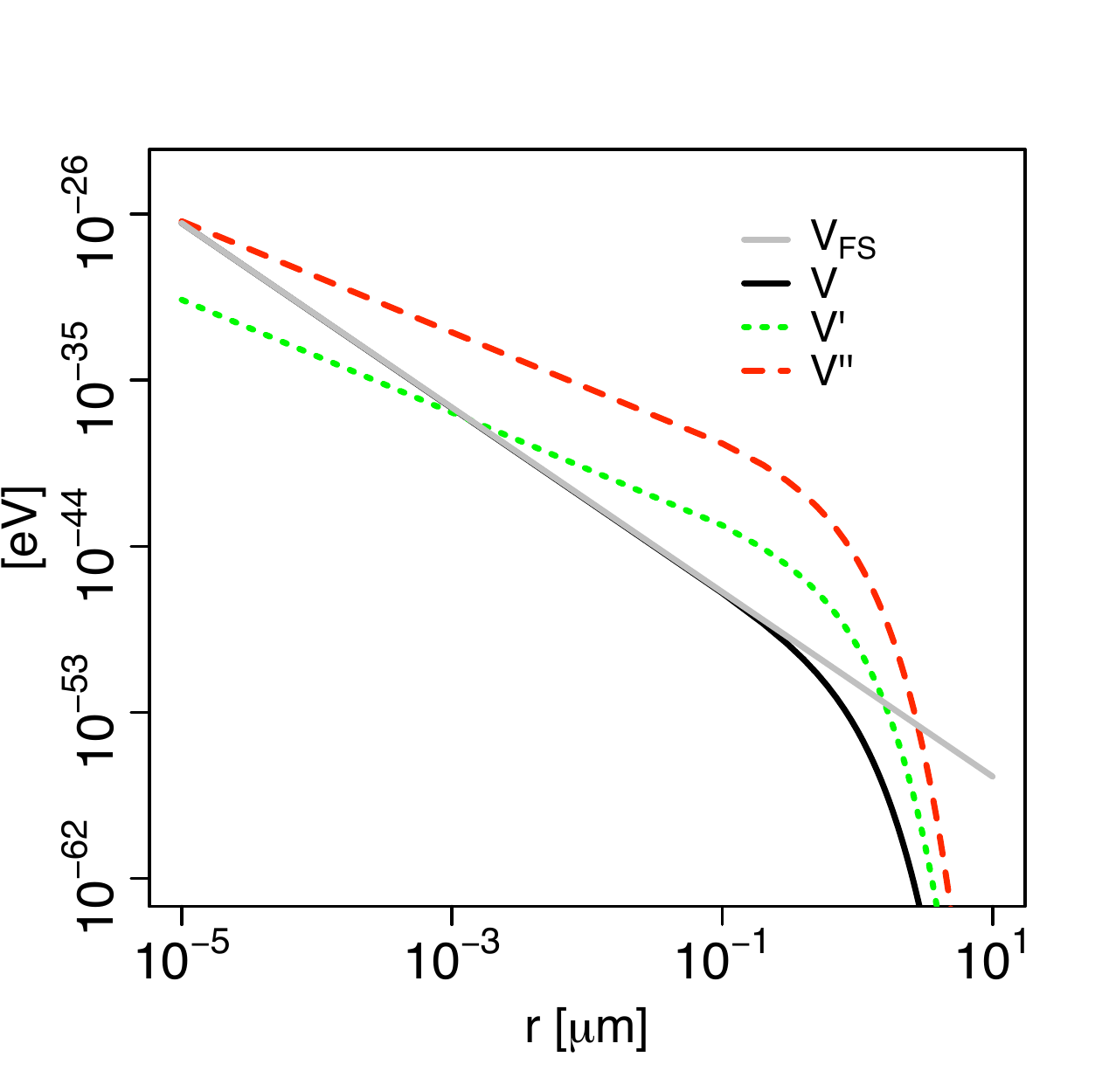}
    \caption{ Potentials (in eV)  as  functions of the distance $r$ (in $\mu$m). The parameters are $m_{\nu}=1$~eV and $\mu_\nu = 10^{-11}  \mu_{\rm Bohr} $.}
    \label{fig:fig3}
   \end{figure}

We did not consider other possible new physics contributions, limiting our attention to neutrino magnetic moments. In this respect, to attempt an interpretation of the anomalous results obtained by LSND, KARMEN and MiniBooNe experiments, there has recently been a suggestion \cite{russo} of exotic neutrinos with much larger magnetic moments: their contribution to the potential $V''(r)$ would be considerable, were it not for the fact that the mass of the new neutrino should be $\aprge$40 MeV, and therefore its contribution would be exponentially suppressed for distances larger than a few femtometers. 

 \section{Effects of neutrino mixing}
 \label{mixing}
 \qquad As we know, flavour eigenstates do not coincide with mass eigenstates. The effect of neutrino mixing and of mass differences should be accounted for. 
 In the scattering amplitude given in eq.(\ref{ampl}),  $K^{\mu \nu}$ defined in eq.(\ref{loop}) should be modified as follows:
 \begin{equation}
 K^{\mu \nu} =  \sum_{i,j=1}^{3} |U_{ei}|^2 |U_{ej}|^2 
  \int {d^4 k \over (2 \pi)^4} {{\rm Tr} [(\widehat{k} +m_i) \gamma^\mu (1- \gamma_5)
 (\widehat{q}+\widehat{k} +m_j) \gamma^\nu (1-\gamma_5)] \over {(k^2-m_i^2)\,((q+k)^2-m_j^2)}}
\label{loop2}
\end{equation}
 As a consequence the long range, spin independent potential becomes
 \begin{eqnarray}
 V(r) = {G_F^2 \over 24\, \pi^3\, r^5} \sum_{i,j=1}^{3} |U_{ei}|^2 |U_{ej}|^2 
  \int_{(m_i+m_j) r}^\infty \sqrt {y^4-2\, (m_i^2+m_j^2)\,r^2 y^2+(m_i^2-m_j^2)^2\,r^4 }
  \nonumber \\ 
  \cdot \left(y^2-{(m_i^2+m_j^2)\, r^2 \over 2} 
 + {(m_i^2-m_j^2)^2\, r^4 \over 2y^2}\right)\;  
 {e^{-y} \over y} \;dy\;,
 \label{potential2}
\end{eqnarray}
 instead of eq.(\ref{potential}). 
 
We only know from oscillation experiments the differences of the neutrino masses squared. 
A nice and up--to--date review of the present situation is given in ref.\cite{Strumia}. 
We will consider three somehow extreme cases as examples.
In the so--called degenerate case the neutrino masses are almost equal and near to the present upper bound from tritium beta decay (and cosmology) of about 1 eV. In this case there is no effect from mixing and the cutoff in the potential due to the neutrino masses occurs for distances larger than 
0.5~eV$^{-1} \sim$ 0.1 $\mu$m. In the case of hierarchical masses we have the two possibilities of ''normal" (i.e. $\nu_1$ is the lightest neutrino, with $m_1 \sim 0$) or ''inverted" (i.e. with $m_1 \sim \sqrt{\Delta_{\rm atm}} \simeq 0.05$ eV and $m_3 \sim 0$). Assuming for the masses and mixing parameters the values $|U_{e1}|=0.84$, $|U_{e2}|=0.54$, 
$\sqrt{\Delta_{\odot}}=0.0087$ eV \cite{Strumia} , in the inverted hierarchical case there is again a single cutoff due to neutrino masses at a distance of  2 $\mu$m $\simeq (2 \sqrt{\Delta_{\rm atm}})^{-1}$, due to the smallness of $U_{e3}$. In the case of normal hierarchy, and assuming vanishing $m_1$ and $U_{e3}$, the above formula (\ref{potential2}) implies that  the potential will be of full strength for distances less than about 11.5 $\mu$m $\simeq (2 \sqrt{\Delta_{\odot}})^{-1}$, its strength will be reduced by about  ten percent up to a distance 23 $\mu$m and it will be halved for  larger distances.

A similar modification should be made for the terms in the potential due to the hypothetical magnetic moment. In the presence of three mass eigenstates of Dirac neutrinos, the magnetic moments will be given by a $3 \times 3$ matrix. The resulting formulae in the general Dirac case are easily obtained, but somehow lengthy; we prefer to consider, as an example,  the more theoretically appealing Majorana neutrinos, in which case the diagonal entries of the neutrino magnetic moment matrix vanish. We further  assume for simplicity a common value $\mu_{\nu}^{\rm Maj}$ for all the transition magnetic moments. The resulting discontinuity is (cfr. eq.(\ref{absKprime})):
\begin{eqnarray}
& & [ K'^{\mu \nu}_{\rm Maj} ] ={- i \over 4 \pi}\; \left(g^{\mu \nu}
- {q^\mu q^\nu \over q^2} \right)\; \sum_{i,j=1}^{3} |U_{ei}|^2 |U_{ej}|^2  \;(1-\delta_{ij})\;
 (m_i+m_j)   \cdot \nonumber \\
 & & \sqrt{1 - 2 \; {(m_i^2+m_j^2) \over q^2} +
{(m_i^2-m_j^2)^2 \over q^4}} \;  \left[1+ {(m_i - m_j)^2 \over q^2}\right]\;\theta[q^2-(m_i+m_j)^2]\; ,
\label{absKprime2}
 \end{eqnarray}
 and therefore the potential becomes:
 \begin{eqnarray}
 V'_{\rm Maj}(r)  =  {{G_F\;\alpha} \over {8 \sqrt{2}\, \pi^2 \, r^3}} \; {1 \over m_e}\; 
 {\mu_{\nu}^{\rm Maj} \over \mu_{\rm Bohr}}\; \sum_{i,j=1}^{3} |U_{ei}|^2 |U_{ej}|^2  \;(1-\delta_{ij})\;
 (m_i+m_j) \cdot \nonumber \\
 \int^\infty_{(m_i+m_j) r} \sqrt{y^2-2\,(m_i^2+m_j^2)\,r^2+{(m_i^2-m_j^2)^2\; r^4 \over  y^2}}\;
  \left[1+ {(m_i - m_j)^2 \;r^2\over y^2}\right]\;e^{-y}\;dy\; .  
 \label{potprime2}
 \end{eqnarray}
 As a consequence of the vanishing diagonal terms in the double sum, the resulting potential is suppressed by a factor $|U_{e1}|^2\;|U_{e2}|^2\simeq 0.4$ with respect to eq.(\ref{potprime}): the cutoff due to the masses is at 23 $\mu$m (2 $\mu$m) for the ''normal" (''inverted") hierarchy, respectively. The degenerate case with masses all equal to $\sim$1 eV is excluded by the null results \cite{0nu2beta} of neutrinoless double beta decay experiments (see, however, \cite{Klapdor}).

\section{Conclusions}
 \qquad In this paper we have evaluated the effect of neutrino masses on the long--range potential due to the exchange of a neutrino-antineutrino pair between two particles carrying weak charge. We considered in particular two electrons. This brings up--to--date calculations present in the literature \cite{Feinberg,Sikivie} for massless  neutrinos. The results are given in Section~\ref{mass}, eq.(\ref{potential}), and Section~\ref{mixing}, eq.(\ref{potential2}),  respectively without and with the effect of mixing, and show that the neutrino masses decrease the potential with respect to the massless case for all distances.
 
In Section~\ref{munu} we considered for the first time the effect of a hypothetical neutrino magnetic moment on the  potential. We noted that these terms are very often dominant over the weak contribution. The effect of mixing has been discussed in Section~\ref{mixing} for one of the terms in a particular simplified case with Majorana neutrinos.

However, for neutral atoms on top of the contribution due to exchange of two--neutrino pairs  between electrons, there will be other contributions due to exchanges between protons and between electrons and protons. As a consequence of the matter neutrality, the contributions of the  terms depending on the neutrino magnetic moment will cancel. Effects related to the nucleon's anomalous magnetic moments could survive, but they are vanishing in the nonrelativistic limit. 
 
 We are indebted to Gianni Carugno for many interesting and stimulating discussions.

\end {document}